# Skyrmions in a magnetic field and $\pi^0$ domain wall formation in dense nuclear matter

Shi Chen, Kenji Fukushima, and Zebin Qiu

*Department of Physics, The University of Tokyo, 7-3-1 Hongo, Bunkyo-ku, Tokyo 113-0033, Japan*

We elucidate magnetic effects in the skyrmion system to probe into the dense nuclear matter. We find a deformed $\pi^0$ dipole structure of a baryon induced by a magnetic field. We then extend our scope to stacked Skyrme crystal layers to scrutinize phases of magnetized nuclear matter. We observe a quantized magnetic flux and identify a phase transition from a crystalline state to a $\pi^0$ domain wall corresponding to a topological transmutation from $\pi_3(S^3)$ to $\pi_1(S^1)$. We establish the phase diagram, which could be explored also in analogous systems with two-component Bose-Einstein condensates.

DOI: 10.1103/PhysRevD.105.L011502

## I. INTRODUCTION

It is important to deepen our understanding of dense baryonic matter based on quantum chromodynamics (QCD). The QCD phase diagram has not been established due to the sign problem at finite density. Intermediate energy heavy-ion collisions (see [1,2] for recent highlights) aim to unveil properties of dense matter and prompt next-generation approaches; see [3–6] for some attempts. Besides, the structure of neutron stars and the dynamics of the neutron star merger crucially depend on the equation of states of dense matter. Numerous endeavors are devoted to capture dense matter properties by the interpolation between the low-density nuclear matter and the high-density quark matter; see [7,8] for reviews.

We could bypass finite-density difficulties with additional parameters such as a magnetic field ***B***. In fact noncentral nuclear collisions (see [9,10]) as well as the neutron star surface accommodate a strong ***B***. The finite-density sign problem at strong ***B*** could be moderate [11], as studied with a two-color QCD benchmark test in [12]. In nuclear physics the state of matter influenced by ***B*** is a fashionable subject. Preceding works include an effect of the magnetic moment [13], a state of quark matter called the chiral magnetic spiral [14] (analogous to inhomogeneous phases realized as a quarkyonic spiral [15,16], the dual chiral density wave [17], the *P*-wave pion condensate [18]), and importantly in the present context, a state of nuclear matter called the chiral soliton lattice (CSL) [19,20].

Because a strong ***B*** makes $\pi^\pm$ gapped, the CSL state is dominated by $\pi^0$ only, and the topologically nontrivial $\pi^0$ winding coupled to ***B*** gives rise to a finite baryon number, which is also the case for the $\pi^0$ domain wall [20] (see also [21]). The system has only U(1) symmetry associated with massless $\pi^0$, and the spatially one-dimensional winding is homotopically characterized by $\pi_1(\mathrm{U}(1))$. In contrast, without ***B***, all $\pi^\pm$ and $\pi^0$ are massless corresponding to SU(2) symmetry, and the baryon number originates from the spatially three-dimensional winding by $\pi_3(\mathrm{SU}(2))$ [22], with which the topological solitons are identified as the baryons [23]. We can summarize our question in a schematic manner as follows:

| Strong-$B$ | $\leftarrow ? \rightarrow$ | Weak-$B$ |
|---|---|---|
| $\pi^0$ Domain-wall | | (Inhomogeneous) Nuclear matter |
| $\pi_1(\mathrm{U}(1))$ winding | | $\pi_3(\mathrm{SU}(2))$ winding |

Two different homotopy groups lie along the *B*-axis but these windings represent the common conserved baryon number. Then, the question is how they could be connected with varying ***B*** and/or density.

To target this concretely we employ the Skyrme model [24–27]. An *ad hoc* interplay between the CSL and the Skyrme crystals was discussed in [28] previously, but we will clarify a self-consistent picture. We also mention related works in various dimensions [29,30]. Here, we would stress the virtues of this work: (1) The Skyrme model is not a phenomenological composition but a minimal extension of the low-energy description of QCD. Most of our results are model independent through a *B* dependent baryon current. (2) We discuss the phase diagram of topological phases belonging to different homotopy groups including the $\pi^0$ domain walls. This is an important step to quantify the relevance of the $\pi^0$ domain walls. (3)

The implementation of adiabatic and continuous deformation of theory connecting different homotopy groups opens a novel research opportunity; namely, topological matter science in nuclear physics. Our arguments apply to similar topological systems in condensed matter physics, e.g., magnetic skyrmions in lower dimensions (see [31] for a review). An even more similar system is found, that is, a three-dimensional two-component Bose-Einstein condensed system [32]. Amazingly, the internal space of the wave function spans over $S^3$, and the three-dimensional skyrmion can be experimentally induced through a vortex ring. The above question is then recast by replacing $\boldsymbol{B}$ with an external disturbance to break two-component symmetry.

## II. SKYRMIONS WITH ELECTROMAGNETISM

Infrared degrees of freedom generated by the spontaneous breaking of chiral symmetry are embodied by $\Sigma = e^{i\boldsymbol{\pi}\cdot\boldsymbol{\tau}/f_\pi} \in \mathrm{SU}(N_f)$ where $N_f$ is the flavor number and $\boldsymbol{\pi}$ represents the Nambu-Goldstone bosons. The infrared amplitude of $\boldsymbol{\pi}$ is reproduced by a quadratic action plus a mass term

$$S_0 = -\frac{f_\pi^2}{4}\int d^4x [\mathrm{tr}(\ell_\mu\ell^\mu) + 2m_\pi^2 \mathrm{tr}(1-\Sigma)], \quad (1)$$

where $\ell_\mu \equiv \Sigma^\dagger D_\mu \Sigma$ with the covariant derivative $D_\mu\Sigma \equiv \partial_\mu\Sigma - iA_\mu[Q,\Sigma]$ for an electric charge $iQ \in \mathfrak{u}(N_f)$. The elementary charge $e$ is absorbed in the gauge field $A_\mu$ in our convention. The 't Hooft anomalies are matched by the Wess-Zumino-Witten action, $S_{\mathrm{WZW}}$ [33,34], i.e.,

$$S_{\mathrm{WZW}} = \frac{i}{80\pi^2}\int d^5x \epsilon^{\bar\mu\bar\nu\bar\rho\bar\sigma\bar\tau} \mathrm{tr}(\bar\ell_{\bar\mu}\bar\ell_{\bar\nu}\bar\ell_{\bar\rho}\bar\ell_{\bar\sigma}\bar\ell_{\bar\tau}), \quad (2)$$

where $\bar\ell_\mu \equiv \Sigma D_\mu \Sigma^\dagger$ was introduced. Thus, $\pi_3(\mathrm{SU}(N_f))$-induced topological symmetry is identified as baryon $\mathrm{U}(1)_B$ [22]. Gauging this $\mathfrak{su}(N_f)$ sector appropriately, the covariant derivative is blind to the $\mathfrak{u}(1)_B$ sector of $Q$, which claims an extra coupling, $S_B = q\int d^4x A_\mu J_B^\mu$ with $q \equiv \frac{N_c}{N_f}\mathrm{tr}Q$ and

$$j_B^\mu = \frac{1}{24\pi^2}\epsilon^{\mu\nu\alpha\beta}\left\{\mathrm{tr}(\ell_\nu\ell_\alpha\ell_\beta) + i\frac{3}{2}F_{\alpha\beta}\mathrm{tr}[Q(\ell_\nu - \bar\ell_\nu)]\right\}. \quad (3)$$

Up to this point there is no model assumption. The Skyrme model assumes the following term to evade Derrick's scaling theorem [35],

$$S_{\mathrm{Sky}} = \frac{1}{32a^2}\int d^4x \mathrm{tr}([\ell_\alpha,\ell_\beta][\ell^\alpha,\ell^\beta]), \quad (4)$$

as the simplest extension. In this sense quantities involving $a$ are model dependent. However, we can scale out the mass dimension by a combination of $f_\pi$ and $a$, and the results in terms of rescaled dimensionless variables should be robust.

We will show results for $m_\pi = 0$ only, but will report the $m_\pi \neq 0$ results in a full length paper [36]. We focus on the $N_f = 2$ case; then $S_{\mathrm{WZW}}$ is simply a $\mathbb{Z}_2$ $\theta$-angle from $\pi_4(\mathrm{SU}(2))$, which is nontrivial for odd $N_c$, agglutinating $\mathrm{U}(1)_B$ to other transformations via the fermionic statistics [22]. For $N_c = 3$, the physical charge $Q = \mathrm{diag}(\frac{2}{3},-\frac{1}{3})$ breaks isospin symmetry into axial isorotation. We introduce a homogeneous downward magnetic field $\boldsymbol{B} = -B\hat{z}$ with $B \geq 0$, adopting $A_0 = 0$ and $\boldsymbol{A} = \frac{1}{2}B\boldsymbol{r}\times\hat{z}$.

## III. SKYRMION DEFORMED BY A MAGNETIC FIELD

We adopt an *axial hedgehog* ansatz for a static skyrmion on account of the pseudoaxial symmetry, $\Sigma(\boldsymbol{r}') = e^{i\alpha Q}\Sigma(\boldsymbol{r})e^{-i\alpha Q}$, where $\boldsymbol{r}'$ denotes the coordinate rotated by $\alpha$ along the $z$-axis. Parametrizing $\Sigma = i\boldsymbol{\tau}\cdot\boldsymbol{\Pi} + \Pi_4$ with

$$\Pi_1 = \sin f \sin g \cos\varphi, \qquad \Pi_3 = \sin f \cos g,$$
$$\Pi_2 = \sin f \sin g \sin\varphi, \qquad \Pi_4 = \cos f, \quad (5)$$

we can express the axial hedgehog as $f = f(r,\theta)$, $g = g(r,\theta)$, $\varphi = \varphi_0(r,\theta) + \phi$ with the spherical coordinates $(r,\theta,\phi)$. The quantization condition of a unit baryon number leads to $f(\infty,\theta) = 0$, $f(0,\theta) = \pi$, $g(r,0) = 0$, $g(r,\pi) = \pi$. En passant, the spherical hedgehog [26] is recovered provided $f = f(r)$, $g = \theta$, and $\varphi = \phi$.

The static Euler-Lagrange equation minimizes the energy functional $M(f,g,\varphi_0)$, i.e., the spatial integral of $T^{00}$, with $T^{\mu\nu}$ the energy-stress tensor. Our ansatz yields,

$$M = 2\pi\int_0^\infty dr r^2 \int_{-1}^1 d\cos\theta T^{00}(r,\theta) \quad (6)$$

with

$$T^{00} = \frac{f_\pi^2}{2}\{|\boldsymbol{\nabla}f|^2 + \sin^2 f[|\boldsymbol{\nabla}g|^2 + \sin^2 g(\Upsilon^2 + |\boldsymbol{\nabla}\varphi_0|^2)]\}$$
$$+ \frac{1}{2a^2}\sin^2 f\{|\boldsymbol{\nabla}f\times\boldsymbol{\nabla}g|^2 + \sin^2 g(\Upsilon^2(|\boldsymbol{\nabla}f|^2 + \sin^2 f|\boldsymbol{\nabla}g|^2)$$
$$+ |\boldsymbol{\nabla}f\times\boldsymbol{\nabla}\varphi_0|^2 + \sin^2 f|\boldsymbol{\nabla}g\times\boldsymbol{\nabla}\varphi_0|^2]\}, \quad (7)$$

where we introduced the following notation,

$$\Upsilon \equiv \frac{1}{r\sin\theta} - \frac{Br\sin\theta}{2}, \quad (8)$$

$$|\boldsymbol{\nabla}h|^2 = (\partial_r h)^2 + \frac{1}{r^2}(\partial_\theta h)^2, \quad (9)$$

$$|\boldsymbol{\nabla}h\times\boldsymbol{\nabla}w|^2 = \frac{1}{r^2}(\partial_r h\partial_\theta w - \partial_\theta h\partial_r w)^2. \quad (10)$$

Here, $h$ and $w$ are either $f$, $g$, or $\varphi_0$.

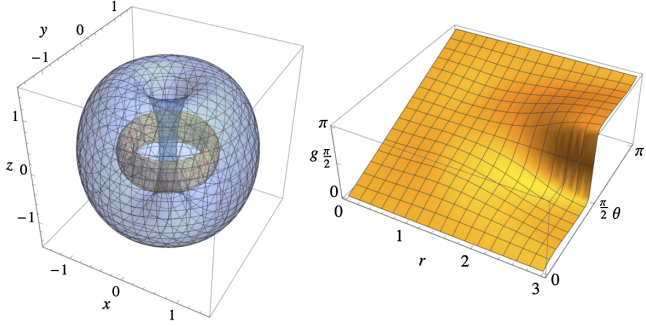

FIG. 1. (Left) Graphical representation of $\pi_3(SU(2))$ with $B = 3.00 f_\pi^2 a^2$; $\Pi_1^2 + \Pi_2^2 = 0.90$ on the inner torus (orange) and $\Pi_3^2 + \Pi_4^2 = 0.90$ on the outer torus (blue). (Right) $g(r,\theta)$ with $B = 3.00 f_\pi^2 a^2$. ($x$, $y$, $z$, and $r$ are of unit $f_\pi^{-1} a^{-1}$.)

Clearly, a uniform $\varphi_0(r,\theta)$ (denoted by $\varphi_0$ hereafter) minimizes the energy functional that is quadratic in $\nabla\varphi_0$. Then employing the finite element method, we solve the Dirichlet problem of $f(r,\theta)$ and $g(r,\theta)$ from $\delta M/\delta f = 0$ and $\delta M/\delta g = 0$ with the boundary conditions prescribed below Eq. (5). Figure 1 shows the spatial profile at $B = 3.00 f_\pi^2 a^2$. The orange inner and the blue outer tori in Fig. 1 (Left) represent surfaces of $\Pi_1^2 + \Pi_2^2 = 0.90$ and $\Pi_3^2 + \Pi_4^2 = 0.90$, respectively. Evidently, the inner torus is laced with a brace from the outer torus, which is a graphical representation of $\pi_3(SU(2))$; see also [32]. Besides, as shown in Fig. 1, our numerical solution turns out to satisfy

$$f(r, \pi - \theta) = f(r, \theta),$$
$$g(r, \pi - \theta) = \pi - g(r, \theta). \quad (11)$$

With a uniform $\varphi_0$, we will refer to these relations by "pseudoreflection symmetry" in this work.

From Eq. (7) we can understand magnetic effects by visualizing a "barrel" of $\Upsilon = 0$, i.e., $r\sin\theta = \sqrt{2/B}$. To minimize $\Upsilon^2 \sin^2 f \sin^2 g = \Upsilon^2 (\Pi_1^2 + \Pi_2^2)$ in the energy functional, nonzero $\Pi_{1,2}$ (charged pions) are localized near the barrel. For $B \lesssim 0.4 f_\pi^2 a^2$, a broad barrel mildly expands the skyrmion profile transversely (i.e., *oblate deformation*). For $B \gtrsim 0.4 f_\pi^2 a^2$, constricted by a narrower barrel, the profile transversely shrinks (i.e., *prolate deformation*), as shown in Fig. 1 (Left). This magnetic scale $\sim 0.4 f_\pi^2 a^2$ reflects the original skyrmion size. For $r$ far from the barrel of $\Upsilon = 0$ the exterior $\Pi_{1,2}$ are suppressed, as shown by the angular distortion of $g(r,\theta)$ in Fig. 1 (Right), because of large $|\Upsilon|$ there. A sharp leap from $g \sim 0$ to $\pi$ near $\theta = \frac{\pi}{2}$ for $r \gtrsim \sqrt{2/B}$ implies that the skyrmion pulls its $\pi^\pm$ cloud in and leaves a sheer $\pi^0$ dipole outside. We shall soon later discuss how transversely contiguous $\pi^0$ dipoles can condense into a $\pi^0$ domain wall.

## IV. DOMAIN WALL FORMATION FROM A SKYRME CRYSTAL

Now we address the $\pi^0$ domain wall formation. As argued in [19] dense nuclear matter under strong $B$ exhibits the CSL, which is approximately viewed as stacked layers of the $\pi^0$ domain walls as illustrated in Fig. 2. Other phase candidates [17,18] also exhibit a multilayer structure on account of the anisotropy induced by ***B***. Hereafter we focus on a single 2D layer for simplicity, and 3D analyses are ongoing [36].

We follow the prescription in [37] to actualize a static 2D Skyrme crystal on a square lattice with pseudoperiodicity that blends crystalline translations with $e^{i\pi I_3}$ (see also [38] for crystallization in a holographic QCD model). In the presence of the vector potential, crystalline translation should incorporate appropriate gauge transformations, i.e., the pseudotranslation symmetry reads,

$$\tau^3 \Sigma(x,y,z) \tau^3 = e^{i\lambda By Q} \Sigma(x + 2\lambda, y, z) e^{-i\lambda By Q},$$
$$\tau^3 \Sigma(x,y,z) \tau^3 = e^{-i\lambda Bx Q} \Sigma(x, y + 2\lambda, z) e^{i\lambda Bx Q}, \quad (12)$$

where we introduced the lattice constant $2\lambda$. We set one baryon in each unit cell. Inspired by the single baryon result, we impose the pseudoreflection symmetry (11) to simplify the problem: We can focus on an octant cell with $0 \le x \le \lambda$, $0 \le y \le \lambda$, and $z \ge 0$, thanks to

$$\tau^1 \Sigma(x) \tau^1 = \Sigma^\dagger(-x),$$
$$\tau^2 \Sigma(y) \tau^2 = \Sigma^\dagger(-y),$$
$$\tau^3 \Sigma(z) \tau^3 = \Sigma^\dagger(-z), \quad (13)$$

which are easily verified from the pseudoreflection symmetry.

The solution in this octant cell is subject to the following boundary conditions. Our vacuum convention is $\Pi_4(x, y, +\infty) = 1$. Equation (11) requires $\Pi_1(0, y, z) = \Pi_2(x, 0, z) = \Pi_3(x, y, 0) = 0$. We can also derive

$$\Pi_1(\lambda, y, z) \sin\left(\frac{1}{2}\lambda By\right) - \Pi_2(\lambda, y, z) \cos\left(\frac{1}{2}\lambda By\right) = 0,$$
$$\Pi_1(x, \lambda, z) \cos\left(\frac{1}{2}\lambda Bx\right) - \Pi_2(x, \lambda, z) \sin\left(\frac{1}{2}\lambda Bx\right) = 0, \quad (14)$$

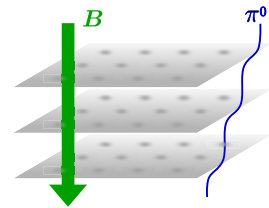

FIG. 2. Illustration of the CSL as approximated by the 2D Skyrme crystal layers.

from Eqs. (11) and (12). We can show that there are two distinct classes of solutions from a condition for the baryon number to be $1/8$ in the octant cell. We recast the baryon number

$$N_B \equiv \int d^3x\, j_B^0 \\ = \frac{1}{4\pi^2}\int (d\varphi - A) \wedge (\Pi_4 d\Pi_3 - \Pi_3 d\Pi_4), \quad (15)$$

as a surface integral. Foregoing conditions force it to vanish except on two edges; that is, we find,

$$N_B = N_0 + N_\lambda, \quad N_0 = \frac{1}{8}n_0, \quad N_\lambda = \frac{n_\lambda}{8}\left(\frac{2}{\pi}\lambda^2 B - 1\right), \quad (16)$$

where $N_0$ appears from an edge at $x = y = 0$ and $N_\lambda$ from another edge at $x = y = \lambda$, with $n_0, n_\lambda \in \mathbb{Z}$ being the $\Pi_{3,4}$ winding number from $z = +\infty$ to $-\infty$ along the according edges [see Fig. 3 (Left)]. We find that $N_B = 1/8$ is to be realized as $N_0 = 1/8$ (with $n_0 = 1$) and $N_\lambda = 0$, leading to two types of crystal solutions; the condition of $N_\lambda = 0$ needs either $n_\lambda = 0$ or $\frac{2}{\pi}\lambda^2 B = 1$ and correspondingly we have additional boundary conditions for $\Pi_4$. Once the boundary conditions are all fixed, we should minimize the energy functional $M(\Pi_{1,2,3,4})/8$ with a Lagrange multiplier that constrains $\sum_{i=1}^4 \Pi_i^2 = 1$. Due to the lack of pseudoaxial symmetry, it is technically convenient to treat $\Pi_i$ rather than $f, g, \varphi$, and the energy functional is then given by

$$\frac{M}{8} = \int d^3x \left(\frac{f_\pi^2}{2}|D\Pi_i|^2 + \frac{1}{4a^2}|D\Pi_i \times D\Pi_j|^2\right), \quad (17)$$

where $D\Pi_1 = \nabla\Pi_1 - A\Pi_2$ and $D\Pi_2 = \nabla\Pi_2 + A\Pi_1$.

First, we refer to the solution with $n_\lambda = 0$ as the *normal crystal*. In this case the boundary conditions, $\Pi_4(0,0,0) = -1$ and $\Pi_4(\lambda,\lambda,0) = 1$, should be satisfied as illustrated in Fig. 3 (Left). On the layer $\Pi_4$ is inhomogeneous, as depicted in the top left corner on the phase diagram in Fig. 5, and we can regard it as a *regional* $\pi^0$ domain wall. In Fig. 4 we plot the crystalline baryon mass $M$ calculated from Eq. (17) as a function of the specific transverse area $\Lambda \equiv 4\lambda^2$ under different $B$. We have confirmed that the isolated baryon mass is recovered asymptotically by $M(\Lambda \to \infty)$. The crystalline structure is stable as seen from the energy minimum whose location is denoted by $\Lambda_0$. We find that the specific binding energy, $M(\infty) - M(\Lambda_0)$, grows monotonically from $6.40 f_\pi a^{-1}$ at $B = 0$ up to a saturated value $\sim 30 f_\pi a^{-1}$ with increasing $B$. Such magnetically facilitated crystallization confirms our $\pi^0$-dipole condensation scenario.

Second, we call the solution with $\frac{2}{\pi}\lambda^2 B = 1$ or $\Lambda B = 2\pi$ the *anomalous crystal*. In this case $N_\lambda = 0$ holds for any $n_\lambda$ and, we find that a transversely uniform $\pi^0$ domain wall with $\Pi_4(x,y,0) = -1$ solves the anomalous crystal. Thus, our anomalous crystal turns out to be a synonym of the conventional *global* $\pi^0$ domain wall [20].

Interestingly, two different boundary conditions, $\Pi_4(\lambda,\lambda,0) = +1$ and $-1$, signify two distinct windings, $\pi_3(SU(2))$ and $\pi_1(U(1))$, respectively. In fact, the normal crystal with inhomogeneous $\Pi_4$ can sustain the $\pi_3(SU(2))$ linkage as shown in Fig. 3 (Right), that is a crystalline version of winding in Fig. 1 (Left), but the anomalous crystal cannot.

## V. THERMODYNAMICS AND PHASE TRANSITION

Figure 4 indicates an optimal $\Lambda$, but in thermodynamics we are rather interested in a problem to locate the phase boundary, $\Lambda_c$, where one topological phase overwhelms the other. We can simply identify the crystalline baryon mass $M$ as the free energy per baryon, which is denoted by $\mathcal{E}$ here to emphasize the thermodynamic nature of our problem. For an $N$-baryon normal crystal, the total Helmholtz free energy $F(T, B, N, N\Lambda) \equiv N\mathcal{E}(T, B, \Lambda)$ yields canonical quantities such as the intralayer pressure and the chemical potential

$$\sigma \equiv -\left(\frac{\partial F}{\partial(N\Lambda)}\right)_{T,B,N} = -\left(\frac{\partial \mathcal{E}}{\partial \Lambda}\right)_{T,B}, \quad (18)$$

$$\mu \equiv \left(\frac{\partial F}{\partial N}\right)_{T,B,N\Lambda} = \mathcal{E} + \sigma\Lambda. \quad (19)$$

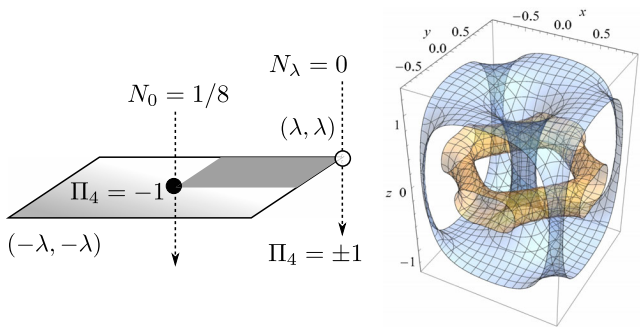

FIG. 3. (Left) $\Pi_{3,4}$ winding contributions to the baryon number contributions $N_0$ at $x = y = 0$ and $N_\lambda$ at $x = y = \lambda$ from the surface of the octant cell (shaded region). (Right) $\pi_3(SU(2))$ in a unit cell with $B = 0.40 f_\pi^2 a^2$ and $1/\Lambda = 0.29 (\gtrsim 1/\Lambda_c)$.

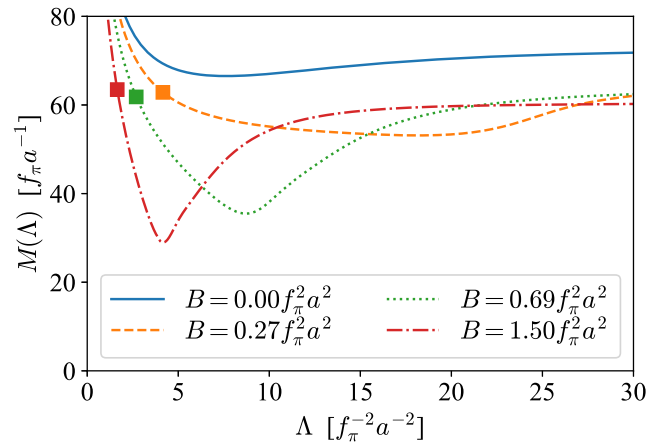

FIG. 4. Crystalline baryon mass $M(\Lambda)$ for various $B$. The square dots denote the location of $\Lambda_c$.

We note that $N\Lambda$ is the 2D volume (area). In contrast, for an $N^*$-baryon anomalous crystal, the constraint by $\Lambda = \Lambda^* = 2\pi/B$ eliminates one degree of freedom, $F^*(T, B, N^*) \equiv N^*\mathcal{E}^*(T, B)$. We use "$*$" for quantities of the anomalous crystal. The anomalous crystal is incompressible at a fixed $B$ and neither $\sigma^*$ nor $\mu^*$ is defined.

At fixed $T$ (where $T = 0$ in the present problem) and $B$, let us consider a mixed system of two crystals and study the phase equilibration by minimizing the total Helmholtz free energy, provided the conservation of the total baryon number and volume (area), i.e.,

$$\delta(F + F^*)|_{T,B,N+N^*,N\Lambda+N^*\Lambda^*} = 0. \quad (20)$$

Specifically, we have $\delta F = \mu \delta N - \sigma \delta(N\Lambda)$ and $\delta F^* = \mathcal{E}^* \delta N^*$. With explicit expressions of $\sigma$ and $\mu$, the equilibrium criterion is

$$\mathcal{E} - \mathcal{E}^* = (\Lambda - \Lambda^*)\mathcal{E}', \quad (21)$$

where $\mathcal{E}'$ is a shorthand for $(\partial \mathcal{E}/\partial \Lambda)_{T,B}$. Let us define $\Lambda_c$ as the critical $\Lambda$ that satisfies Eq. (21). To evaluate $\Lambda_c$ concretely, we note that $\mathcal{E}^* = 16\pi f_\pi^2 m_\pi/B = 0$ [20] in the chiral limit and $\mathcal{E}(\Lambda) = M(\Lambda)$ was presented in Fig. 4. The shape of $M(\Lambda)$-curves, with a repulsive core at small $\Lambda$, ensures the existence of $\Lambda_c$ for any $B > 0$. In Fig. 4, we pinpoint $\Lambda_c$ by a square dot on each $M(\Lambda)$-curve.

In Fig. 5 we plot $\Lambda_c(B)$ and $\Lambda^* = 2\pi/B$, that serves as a phase diagram on the plane of the magnetic field vs the baryon density. The normal crystal manifests itself as a dense high-pressure phase for $\Lambda^{-1} \geq \Lambda_c^{-1}$. For a better intuitive picture we overlay a 2D baryon density profile (at a fixed $B$ and $\Lambda$) in the normal crystal region in Fig. 5. The anomalous crystal appears only on the line of $\Lambda^*$. These two topological phases characterized by $\pi_3(SU(2))$ and $\pi_1(U(1))$ are separated by a first-order phase transition. We have numerically verified that the specific latent heat, $\mathcal{E}(\Lambda_c) - \mathcal{E}^*$, remains finite with a minimum $61.8 f_\pi a^{-1}$ at $B = 0.69 f_\pi^2 a^2$ (the green dot in Fig. 4). We will report more details in a separate publication [36].

## VI. SUMMARY AND OUTLOOKS

We investigated an isolated baryon under $\boldsymbol{B}$ using the Skyrme model, revealing an elliptic deformation. We formulated two 2D Skyrme crystals; a *normal crystal* realizes a regional $\pi^0$ domain wall, while an *anomalous crystal* exhibits a uniform $\pi^0$ domain wall. Interestingly, the former and the latter accommodate the quantized baryon number differently from $\pi_3(SU(2))$ and $\pi_1(U(1))$, respectively. We established a thermodynamic criterion for crystalline transitions and explored the phase diagram with varying magnetic strength and baryon density.

Although our present system was a 2D layer, stacked normal crystals incarnate inhomogeneous nuclear matter and stacked anomalous crystals realize the CSL. We built a framework in terms of not the chemical potential but an anomalous $B$-term in the baryon current, which is model-independently protected by the anomaly. Our 2D analysis implies that the CSL phase is thermodynamically favored at lower density and stronger $B$, and the ordinary nuclear matter would take over the ground state at higher density and lower $B$. We should quantize the skyrmion to quantify the isospin dependence, to find an intriguing formula for the proton-neutron mass splitting, which will be reported together with 3D analyses in a forthcoming publication. [36].

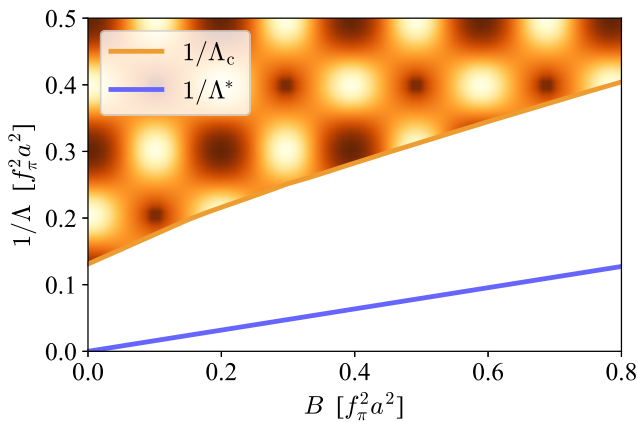

FIG. 5. $\Lambda_c$ and $\Lambda^* = 2\pi/B$ as functions of $B$. In the top left region above $1/\Lambda_c$ the normal crystal is favored, where a 2D density profile is overlaid. The anomalous crystal appears only on the bottom right solid line of $1/\Lambda^*$. We note that $1/\Lambda$ corresponds to the 2D density, and this figure can be regarded as a phase diagram on the $B$-density plane.

## ACKNOWLEDGMENTS

The authors thank Xu-Guang Huang, Muneto Nitta, Mannque Rho, and Naoki Yamamoto for valuable comments and discussions. This work was supported by Japan Society for the Promotion of Science (JSPS) KAKENHI Grant No. 21J20877 (S. C.), No. 18H01211, No. 19K21874 (K. F.), and No. 20J20974 (Z. Q.).